\documentclass{Interspeech}
\interspeechcameraready
\usepackage[dvipsnames]{xcolor}
\usepackage{bbm}

\title{Unlocking Temporal Flexibility: Neural Speech Codec\\ with Variable Frame Rate}

\author[affiliation={1}]{Hanglei}{Zhang}
\author[affiliation={1}]{Yiwei}{Guo}
\author[affiliation={1}]{Zhihan}{Li}
\author[affiliation={2}]{Xiang}{Hao}
\author[affiliation={1}]{Xie}{Chen}
\author[affiliation={1}]{Kai}{Yu}

\affiliation{MoE Key Lab of Artificial Intelligence, Jiangsu Key Lab of Language Computing, X-LANCE Lab, School of Computer Science}{Shanghai Jiao Tong University}{China}
\affiliation{Department of Data Science and Artificial Intelligence}{The Hong Kong Polytechnic University}{China}
\email{op.131@sjtu.edu.cn, kai.yu@sjtu.edu.cn}

\usepackage{comment}
\usepackage{makecell}   % 单元格换行
\usepackage{multirow} 
\usepackage{subfig}
\begin{document}

\maketitle

% the abstract here must exactly match the abstract entered into the paper submission system
\begin{abstract}
Most neural speech codecs achieve bitrate adjustment through intra-frame mechanisms, such as codebook dropout,  at a Constant Frame Rate (CFR).
However, speech segments inherently have time-varying information density (e.g., silent intervals versus voiced regions).
This property makes CFR not optimal in terms of bitrate and token sequence length, hindering efficiency in real-time applications.
% This paradigm forces segments with varying information density (e.g., silent intervals versus voiced regions) to be processed at an identical temporal resolution, hindering minimal bitrate usage and shorter token sequences.
% This is particularly problematic for real-time applications.
In this work, we propose a Temporally Flexible Coding (TFC) technique, introducing variable frame rate (VFR) into neural speech codecs for the first time. 
TFC enables seamlessly tunable average frame rates and dynamically allocates frame rates based on temporal entropy. 
Experimental results show that a codec with TFC achieves optimal reconstruction quality with high flexibility, and maintains competitive performance even at lower frame rates. 
Our approach is promising for the integration with other efforts to develop low-frame-rate neural speech codecs for more efficient downstream tasks.
\end{abstract}

\section{Introduction}
{\let\thefootnote\relax\footnotetext{\vspace{-0.1in}Kai Yu is the corresponding author.}}
Recent audio/speech codecs have integrated neural networks to achieve high-fidelity audio reconstruction~\cite{soundstream,encodec,dac}. 
These models are typically trained end-to-end and consist of three main components: an encoder that compresses the input signal into compact representations, a quantization module
% —often based on Vector Quantization (VQ)\cite{vq}, typically employing Residual Vector Quantization (RVQ)\cite{soundstream}—
that discretizes these representations, and a decoder that reconstructs the audio from the quantized vectors. 
Owing to their success, discrete token-based generative modeling has been successfully extended to the speech domain, leading to speech generation models such as AudioLM~\cite{audiolm} and VALL-E~\cite{valle}.

However, existing neural speech codecs generate token sequences at a high temporal rate—up to 75 frames per second for systems such as EnCodec~\cite{encodec}—whereas text tokenization using BPE~\cite{bpe} typically yields only 3–5 tokens per second. 
This significant discrepancy in token lengths results in inefficient autoregressive prediction, causing performance degradation and increased latency in real-time applications~\cite{whyspeech}. 
In addition to efforts aimed at improving the efficiency of downstream speech generation models themselves~\cite{chen2024valle2neuralcodec,song24b_interspeech,vadusa}, several recent studies have begun to explore low-frame-rate neural audio codecs to fundamentally address this challenge~\cite{lfsc,kyutai2024moshi,semanticodec,ji2024wavtokenizer,stablecodec}.

Despite these efforts, 
% Motivated by the need to reduce sequence lengths, 
we also observe that most neural speech codecs operate as Constant Bitrate (CBR) systems, which could lead to redundant coding, especially for segments like silence where a lower coding rate would suffice compared to highly dynamic areas.
This temporal redundancy undermines the goal of minimizing both overall bitrate consumption and sequence length.  
While prior work~\cite{vrvq} has preliminarily incorporated a Variable Bitrate (VBR) strategy based on information density, it still belongs to the Constant Frame Rate (CFR) framework and does not reduce the overall sequence length.  
To the best of our knowledge, a Variable Frame Rate (VFR) strategy has not yet been introduced to the neural speech codec field.  
% As VFR itself can be integrated orthogonally with other techniques to reduce sequence length and overall bitrate —potentially yielding synergistic benefits that surpass the sum of individual improvements—the exploration space is both broad and promising. 

% We anticipate that a VFR strategy can be integrated orthogonally with other sequence reduction techniques, allowing for a synergistic combination where the overall improvement is greater than the sum of its parts. 
% And it should be designed to be highly transferable, enabling it to be incorporated into various codec backbones with relative ease.
Accordingly, we propose \textbf{Temporally Flexible Coding (TFC)} for neural speech codecs with two primary objectives: (1) to shorten the overall sequence lengths produced at a given total bitrate, and (2) to enable dynamic frame rate allocation that achieves better temporal compactness and flexibility.

TFC introduces VFR coding to neural speech codecs by dynamically adjusting output codes' receptive fields along the temporal axis. 
 Inspired by techniques from variable-rate image compression~\cite{controlgic} and early analyses of speech information entropy in the VFR context~\cite{you2004entropy}, we leverage a non-parametric entropy calculation technique on speech segments to gauge their information density for dynamic frame rate allocation. 
 High-entropy regions, which indicate information-rich content, are allocated higher frame rates, while low-entropy regions, such as silence, are assigned lower frame rates. 
In summary, TFC enables \textbf{a single model to support a user-specified average frame rate ranging from 18.75 Hz to 75 Hz } (following DAC~\cite{dac} backbone for 24kHz audio), and \textbf{ensures a reasonable distribution of time-varying frame rates under a specific total bitrate}.

Overall, TFC achieves both flexible frame rate control and interpretable granularity allocation within a unified framework.
It presents a novel contribution to speech coding by introducing VFR to neural speech codecs for the first time.
% As TFC itself can be integrated orthogonally with other techniques to reduce sequence length and overall bitrate which can potentially yield synergistic benefits that surpass the sum of individual improvements, it shows broad and promising application values. 
Furthermore, the design of TFC can be orthogonally integrated with other techniques aimed at reducing sequence length and overall bitrate, demonstrating broad and promising application prospects.
% making the development of low-bitrate codecs more promising in the future.
% it has the potential to yield synergistic benefits that surpass the sum of its individual improvements, thereby demonstrating broad and promising application prospects.

% In the following sections, we will present background with more  present the architecture of our proposed method with details of the entropy-based allocation algorithm, and evaluate its performance against state-of-the-art benchmarks. Through this work, we aim to advance the field of audio compression and expand its potential applications in generative and multimodal AI systems.
\vspace{-3pt}
\section{Background}
This section situates our work within the broader context of development in feature compression, providing a comprehensive discussion of terms CBR, VBR, CFR, and VFR, as well as the necessary notations for further understanding.
\subsection{Constant versus Variable Bitrate in Neural Codecs}

The concepts of \textbf{Constant Bitrate (CBR)} and \textbf{Variable Bitrate (VBR)} in audio/speech codecs are defined in the temporal domain.
In CBR, the codec maintains a fixed number of bits per unit time, whereas VBR  enables dynamically adjusting the temporal bit allocation based on the complexity of the signal.

Formally, let $\mathbf z_e= f_\theta(\mathbf x)\in \mathbb{R}^{D\times T}$ denote the continuous latent representations of audio $\mathbf x$, encoded by $f_\theta(\cdot)$, where $T$ represents the number of downsampled temporal frames, and $D$ is the dimensionality of the latent vectors.
At a specific frame $t$, the latent vector $\mathbf  z_e[t]$ is quantized to obtain $\mathbf z_q[t]$. 
In an RVQ-based codec with $N_q$ quantizers, $\mathbf z_q[t]$ is computed as 
% \begin{equation}
$\mathbf z_q[t] = \sum_{i=1}^{N_q}Q_i(\mathbf r_i[t])$
% \end{equation}
where $Q_i(\cdot)$ is the $i$-th quantizer, and the residual vector $\mathbf r_i[t]$ is defined recursively as:
\begin{equation}
\mathbf r_i[t] =
\begin{cases}
\mathbf z_e[t] & i = 1, \\
\mathbf r_{i-1}[t] - Q_{i-1}(\mathbf r_{i-1}[t]) & 2 \leq i \leq N_q. \label{eq:rvq-residual}
\end{cases}  
\end{equation}
The reconstructed audio is then obtained as $\mathbf{\hat{x}}=g_\psi(\mathbf z_q)$ using the decoder $g_\psi(\cdot)$.

The quantizer dropout strategy introduced in~\cite{soundstream} allows a single RVQ-based codec to operate at multiple target bitrates.
% without requiring separate training runs. 
Specifically, by randomly sampling $n_q \sim \{1,\cdots,N_q\}$ and using only the first $n_q$ quantizers for different samples during training, the user can flexibly select $n_q$ at inference time, where a lower $n_q$ results in lower reconstruction quality but reduced bitrate usage.
However, this trick alone does not change the fact that RVQ-based codecs allocate the same number of codebooks across all temporal frames under a fixed frame rate, making them effectively CBR.
The needs in generative speech modeling have spurred new concepts for neural codecs, such as single-codebook~\cite{singlecodec,ji2024wavtokenizer,lscodec}, multi-resolutions~\cite{snac,guo2024speaking}, and low-frame-rate or low-bitrate codecs~\cite{lfsc,semanticodec,stablecodec,wu2024ts3}. 
Although these designs empower downstream real-time dialogue applications~\cite{kyutai2024moshi,Xie2024MiniOmniLM}, they remain within the CBR paradigm.

Recently, VRVQ~\cite{vrvq} introduced the VBR strategy into RVQ-based neural audio codecs. 
It assigns a time-varying $n_q$ to different downsampled frames based on a neural-predicted importance map at inference time. 
The design of VRVQ aligns well with efforts to reduce the overall bitrate consumption, but it does not reduce the number of frames.
% For instance, LFSC employs an improved quantization method to yield low-frame rate codes (21.5Hz) while maintaining competitive reconstruction quality, and SNAC introduces an RVQ approach with multi-scale frame rates (12Hz–23Hz–47Hz). 
\begin{figure*}[t]
\vspace{-8pt}
    \centering
    % 右侧大图
    \subfloat[]{
        \includegraphics[width=0.75\linewidth]{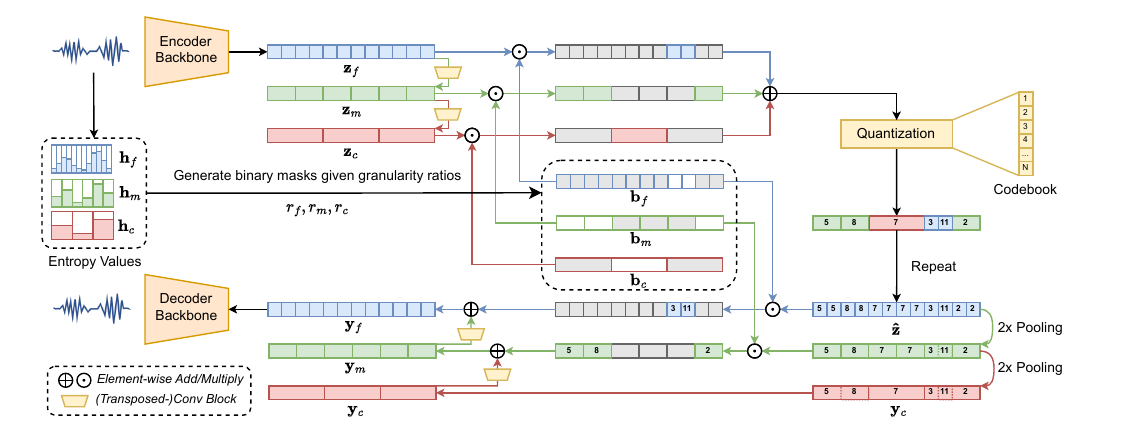}
        \label{TFC}
    }
    % 左侧小图
    \subfloat[]{
        \includegraphics[width=0.25\linewidth]{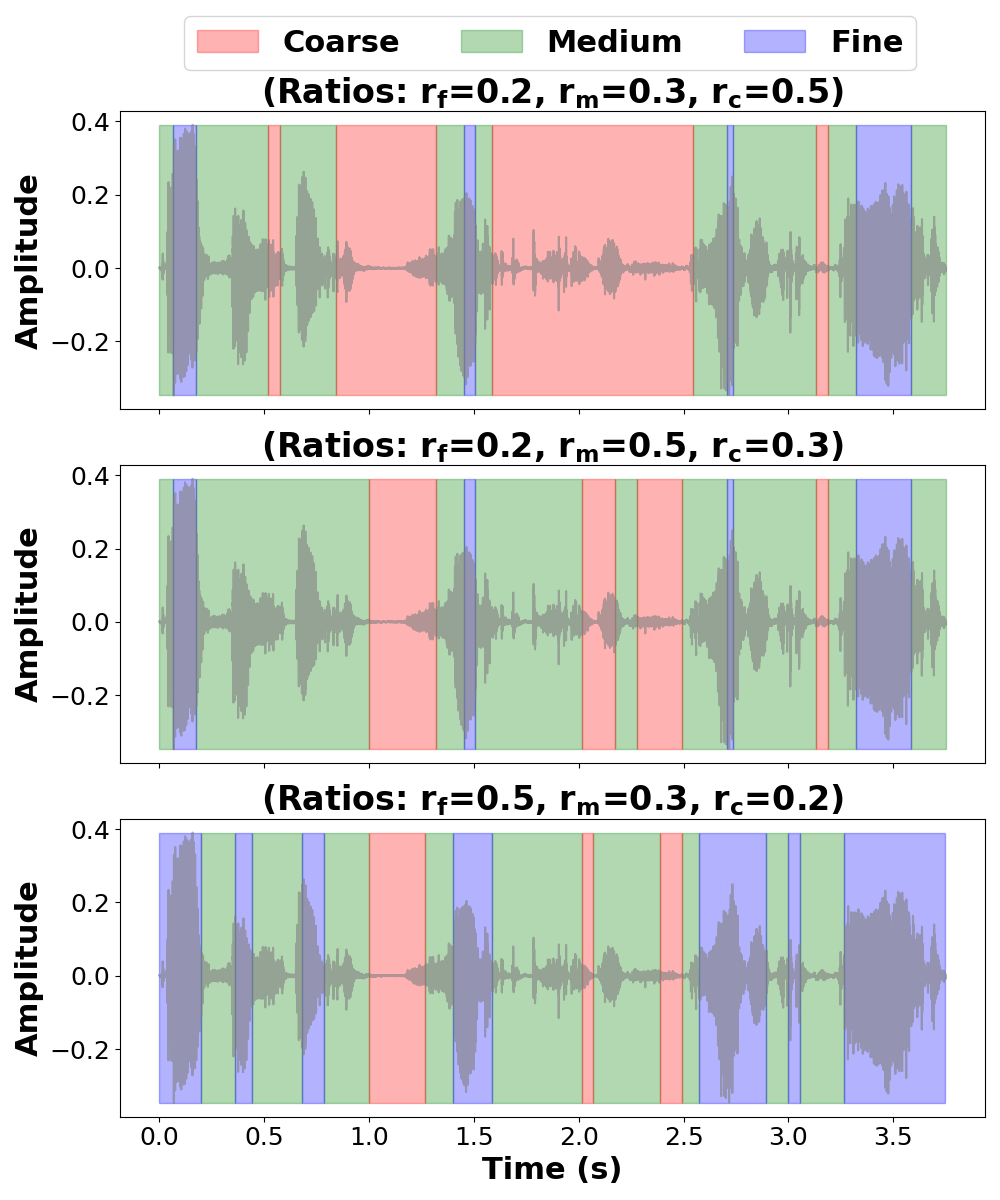}
        \label{allocation}
    }
    \vspace{-0.1in}
    \caption{(a) Proposed framework of Temporally Flexible Coding. (b) VFR allocation examples under different granularity ratios.}
    \vspace{-8pt} % 进一步减少 caption 和正文的距离
    \label{fig:multi_images}
\end{figure*}

\subsection{Variable Frame Rate Compression}
Although achieving VBR coding, VRVQ still operates at a \textbf{Constant Frame Rate (CFR)}, meaning that each encoded frame represents a fixed temporal window of the signal. 
Low frame rates are particularly important for autoregressive modeling, motivating the need to clarify the broader concept of VBR and to discuss \textbf{Variable Frame Rate (VFR)} codecs. 

In the context of audio compression, VBR does not necessarily imply VFR, especially in RVQ-based codecs where each frame can contain multiple quantizers. 
Here, we define VFR as a coding strategy that  \textbf{allows each encoded frame to adaptively cover different temporal spans}. 
It indicates that the model can effectively adjust the downsampling rate based on information density.
For example, a code in VFR setting can cover a longer time span during silent periods and a shorter span during rapidly-changing phoneme transitions.

The VFR concept has been implemented in semantic tokens derived from speech self-supervised learning models~\cite{mohamed2022self}, which are primarily designed for discriminative tasks such as ASR.
For these tokens, preserving fine-grained acoustic details is less critical, and VFR compression can be achieved through distillation from pretrained models or unit discovery techniques~\cite{cho2024sylber,baade2024syllablelm}. 
% For semantic tokens from speech self-supervised learning models~\cite{hsu2021hubert}, which are primarily designed for discriminative tasks such as ASR—where preserving fine-grained acoustic details is less critical—VFR compression can be achieved through distillation from pretrained models or unit discovery techniques~\cite{cho2024sylber,baade2024syllablelm}. 
However, acoustic tokens from neural codecs inherently encode more detailed information; hence, assigning different coding granularities becomes less straightforward due to the absence of explicit supervisory labels that are available for semantic tokens. 
This necessitates the development of effective heuristic strategies for variable frame rate allocation. 
Inspired from variable rate coding methods in image compression\cite{dqvae,controlgic} that adjust spatial resolution based on content complexity, we introduce the VFR strategy to speech codecs to adaptively manage the temporal resolution.
% Along with codebook dropout trick and aforementioned advancements in reducing the overall bitrate for audio coding~\cite{lfsc,wavtokenizer}, the incorporation of VFR methods could further enhance compression efficiency, leading to more flexible bitrate control in audio reconstruction and facilitate the development of more compact representations for future speech-language modeling.

% \begin{figure*}[t]
% \centering
% \includegraphics[width=14cm]{TFC2.jpg}
% \caption{Our TFC}
% \label{fig::model}
% \end{figure*}
\section{Temporally Flexible Coding}
In this section, we will describe our proposed TFC strategy. 
% Since it doesn't introduce additional losses, it's easy to be integrated into codec backbones (e.g., encoder/decoder in DAC).
TFC is a plug-and-play module that can be integrated to various codec backbones like DAC~\cite{dac}, without introducing additional losses for optimization.

\subsection{Measuring Information Density by Temporal Entropy}
\label{sec:entropy-calculation}
% As neural-based entropy estimation risks instability as reported in \footnote{\url{https://github.com/CrossmodalGroup/DynamicVectorQuantization}}, 
% we choose to implement a non-parametric entropy calculation for VFR allocation. 
To measure the information density of a speech segment for granularity allocation in VFR coding, we choose a non-parametric entropy-based approach.
Note that an alternative is to use a neural router for granularity allocation~\cite{dqvae}, but this introduces the risk of instability in gradient estimation, and is reported to perform worse than entropy-based methods in~\cite{dqvae}'s implementation.
Thus we make adaptations upon the algorithm  originally designed for spatial entropy~\cite{spatial,controlgic} to fit the temporal information in speech signals.

Assume $\mathcal T$ is a temporal segment, and $\mathbf x[t]\in[-1,1]$ represents the amplitude value of normalized speech signal $\mathbf x$ at $t\in \mathcal T$. 
% Assume $x\in \mathcal{F}$ denotes an sampling point in a normalized speech segment $\mathcal{F}$, with value $p_x$ in $[-1,1]$. 
We define $N$ uniformly spaced bins $\{u_1,u_2,...,u_N\}$ spanning $[-1, 1]$ as $u_i=-1+\frac{2i}{N-1}$ for $i=0,1,\cdots,N-1$. To estimate the probability density of $\mathbf x[t]$ across these bins, we compute its Gaussian affinity $p_{t,i}$ to each bin $u_i$:  
\begin{equation}
    p_{t,i} = \exp\bigg (-\frac{(\mathbf x[t]-u_i)^2}{2\sigma^2} \bigg ),
\end{equation}
where $\sigma$ controls distribution sharpness. This models the likelihood of $\mathbf x[t]$ ``diffusing" to bin $u_i$.

For a segment $\mathcal{T}$, we average affinities across all $|\mathcal T|$ samples in $\mathcal{T}$ and normalize them to form a probability distribution:
\begin{equation}
    p_{\mathcal{T},i}=\frac{1}{|\mathcal T|}\sum_{t\in\mathcal{T}}p_{t,i}, \quad  \overline{p_{\mathcal{T},i}}=\frac{p_{\mathcal{T},i}}{\sum_{j=0}^{N-1}p_{\mathcal{T},j}+\epsilon},
\end{equation}
where $\epsilon$ ensures numerical stability. The temporal entropy of $\mathcal{T}$ is computed with Shannon Entropy:
\begin{equation}
    H(\mathcal{T}) = -\sum_{i=0}^{N-1} \overline{p_{\mathcal{T},i}} \log \overline{p_{\mathcal{T},i}}
\end{equation}
The computed entropy values, indicating the information densities in specific temporal windows $\mathcal T$, are used for dynamically allocating frame rates described in next section. 
% Then, frame rate allocation can be conducted similarly as spatial resolution allocation. The features can hierarchically  with one code vector $\mathcal{F}$ attends to $w_f=W$ sampling points with hop-size $h$, we denote its frame rate as fine frame rate $f$, in our implementation, we'll have corresponding subsampled versions of code sequences with medium and coarse frame rates $f/2$ and $f/4$, attending to $w_m=W+h$ and $w_c=W+3h$ sampling points respectively. These three groups of 

\subsection{Encoder with Variable Frame Rate Allocation}
\label{encode}
 Let the codec encoder backbone output representation sequence $\mathbf z_f$ with a receptive field $w_f=W$ sampling points and stride $s$. 
 We denote its frame rate as $F$, corresponding to \textbf{fine-resolution representations} $\mathbf z_f$. 
We then generate two additional subsampled representation sequences of $\mathbf z_f$ by CNN blocks:
\begin{itemize}
    \item \textbf{Medium-resolution representations} $\mathbf z_m$, downsampled from $\mathbf z_f$ with frame rate $F/2$, stride $2s$ and each frame's receptive field $w_m=W+s$ sampling points.
    \item \textbf{Coarse-resolution representation} $\mathbf z_c$, downsampled from $\mathbf z_m$ with frame rate $F/4$, stride $4s$ and each frame's receptive field $w_c=W+3s$ sampling points.
\end{itemize}
% To facilitate variable frame rate representations, the feature's resolutions for different temporal areas are allocated based on entropy values. 
% Three groups of entropy values with same lengths as $\mathbf z_f,\mathbf z_m,\mathbf z_c$, respectively, are computed with each resolution's receptive field and stride. 
% Given user-defined granularity ratios $r_0,r_1,r_2$ that satisfy $r_i\in[0,1], r_0+r_1+r_2=1$ and initially zeroed masks $m_0,m_1,m_2$ with same shapes as three entropy groups, we assign 1s to positions in $m_2$ which have lowest $r_2\%$ entropy values. 
% From the remaining medium-resolution entropy values (excluding positions already allocated in $m_2$), we assign 1s to positions in $m_1$ with lowest $r_1\%$ entropy values. 
% The remaining positions, which are 0s in $m_1$ and $m_2$, are then set to 1s in $m_0$. 
For each resolution, we compute entropy values based on Section \ref{sec:entropy-calculation}, where the size of $\mathcal T$ is set to the corresponding receptive field, and temporal segments are slid by the corresponding stride.
Denote $\mathbf h_f,\mathbf h_m,\mathbf h_c$ as the scalar entropy sequences for fine, medium and coarse resolution, respectively, each conforming to the frame rate of $\mathbf z_f,\mathbf z_m,\mathbf z_f$.

Then, we combine the three resolutions together to form a VFR representation.
Given user-defined non-negative scalar granularity ratios $r_f,r_m,r_c$ with $r_f+r_m+r_c=1$, we firstly generate binary sequential granularity masks $\mathbf b_c,\mathbf b_m,\mathbf b_f$ by:
\vspace{-2pt}
\begin{equation}
\begin{aligned}
    \mathbf b_c &= \mathbbm{1} \left( \mathbf h_c \leq \text{Quantile}_{r_c}(\mathbf h_c) \right) \\
    \mathbf h_{m}' &= \mathbf h_m \odot (\mathbf 1 - (\mathbf b_c)\uparrow_2) \\
    \mathbf b_m &= \mathbbm{1} \left( \mathbf h_{m}'\leq \text{Quantile}_{\frac{r_m}{1 - r_c}}\left(\mathbf h_{m}'\right) \right) \\
    \mathbf b_f &= \mathbf 1 - (\mathbf b_m) \uparrow_2 - (\mathbf b_c) \uparrow_4
\end{aligned}
\end{equation}
where $\mathbbm{1}(\cdot)$ is the indicator function, and $\text{Quantile}_q(\mathbf h)$ represents the $q$-quantile of $\mathbf h$ within each utterance. 
$(\cdot)\uparrow_{k}$ denotes repeating operations by factor $k$.
$\mathbf h_{m}'$ represents the remaining part in $\mathbf h_m$ that is not masked by $\mathbf b_c$.
The binary masks $\mathbf b$ have the same frame rate as that of the corresponding $\mathbf h$ and $\mathbf z$ for each resolution.
As a result, a 1-second speech segment will yield $Fr_c/4$ coarse-grained frames, $Fr_m/2$ medium-grained frames and $Fr_f$ fine-grained frames before quantization, whose temporal spans follow a $r_c{:}r_m{:}r_f$ ratio.
% \textcolor{cyan}{If this design is adopted, explanation of $\uparrow$ denotation below needs to be moved here together.}

% \textcolor{red}{TODO: this paragraph may be difficult to understand}

Subsequently, for each resolution, the continuous representations are quantized to produce $\mathbf {\hat{z}}_f,\mathbf {\hat{z}}_m,\mathbf {\hat{z}}_c$, using the RVQ process described in \eqref{eq:rvq-residual}.
To fuse these representations into a unified feature sequence $\mathbf {\hat{z}}$ that aligns with the finest temporal resolution, we repeat the medium and coarse quantized vectors, and integrate them via element-wise multiplication $\odot$ with their corresponding masks:
\begin{equation}
\mathbf {\hat{z}} = (\mathbf {\hat{z}}_f \odot \mathbf b_f) + \left(\mathbf {\hat{z}}_m \odot \mathbf b_m\right) \uparrow_{2} + \left(\mathbf {\hat{z}}_c \odot \mathbf b_c\right) \uparrow_{4}
\end{equation}
% Here, $(\cdot)\uparrow_{k}$ denote repeating operations by factor $k$.
\vspace{-6pt} 

Generally, this hierarchical design allows multiple temporal resolutions to be represented with one frame of codebook indexes, reducing the encoded bitrate for low-entropy area by a smaller frame rate. 
Figure~\ref{allocation} exemplifies this allocation strategy, where we can find clearly that coarse frames are mostly allocated to silent segments, and granularity ratios can be conveniently adjusted for different frame rates.

\subsection{Decoder with Conditional Hierarchical Design}
\label{decode}
Inspired by \cite{controlgic}, we adopt a conditional hierarchical design in decoder to progressively refine the fused latent representations after quantization by gradually integrating information from coarser to finer scales. 
Given the binary granularity masks \( \{ \mathbf b_f, \mathbf b_m, \mathbf b_c \} \), the decoder processes the latent representation $\mathbf {\hat{z}}$ as follows:
\begin{itemize}
    \item The coarse representation $\mathbf y_c$ is derived by downscaling $\mathbf {\hat{z}}$ via average pooling: $\mathbf y_c = \left( \mathbf {\hat{z}} \right) \downarrow_4$.
    \item The medium representation $\mathbf y_m$ is conditioned on $\mathbf y_c$, combined with original quantized medium-resolution latent features.
    Specifically, we use a transposed CNN block to upsample $\mathbf y_c$ by a factor of 2, and add the upsampled sequence with $ \left( \mathbf {\hat{z}} \right) \downarrow_2 \odot \mathbf b_m $.
   \item The finest representation \( \mathbf y_f \) integrates both $\mathbf y_m$ and $\mathbf y_c$, fused with original quantized fine-resolution latent  $\mathbf {\hat{z}} \odot \mathbf b_f $. Similarly, this is also achieved via upsampling $\mathbf y_m$ using a transposed CNN, before adding with $\mathbf {\hat{z}} \odot \mathbf b_f $.
   % The finest representation sequence $\mathbf y_f$ resides in the original frame rate $F$, and is fed to the codec decoder backbone for reconstruction.
\end{itemize}
Overall, each layer refines its predecessor’s output by aggregating reconstructed features with earliest quantized features, avoiding potential error propagation when solely relying on former reconstructed results. Finally, $\mathbf y_c$ with original frame rate $F$ is passed to codec decoder backbone  to recover waveform.

\vspace{-0.08in}
\section{Experiments}
\vspace{-0.02in}
\subsection{Architecture and Setup}
We implement TFC upon the DAC~\cite{dac} backbone, which achieves high-fidelity reconstruction performance among various existing codecs~\cite{wu-etal-2024-codec}. 
We use its official configuration for 24kHz audio that produces RVQ codes at 75Hz frame rate.
Hence in our DAC+TFC framework, the finest granularity is $F=$75Hz.
Our hierarchical design described in Section \ref{encode} performs 2 times $2\times$ downsampling with CNN blocks to obtain medium and coarse representations, and upsamplings as in Section \ref{decode} with transposed CNN blocks reversely. 
The encoded representations are quantized using 8-dimensional RVQ with 1024 (10 bits) entries and maximum codebook number $N_q=32$. 
As this results in a large bitrate (24kbps) that is not practical for downstream generative applications, we take advantage of codebook dropout to use at most 8 codebooks (6kbps) in inference stage. 
\begin{figure*}[ht]
    \centering
    \subfloat[Results for inferences under average 3kbps/37.5Hz.]{
        \includegraphics[width=0.63\linewidth]{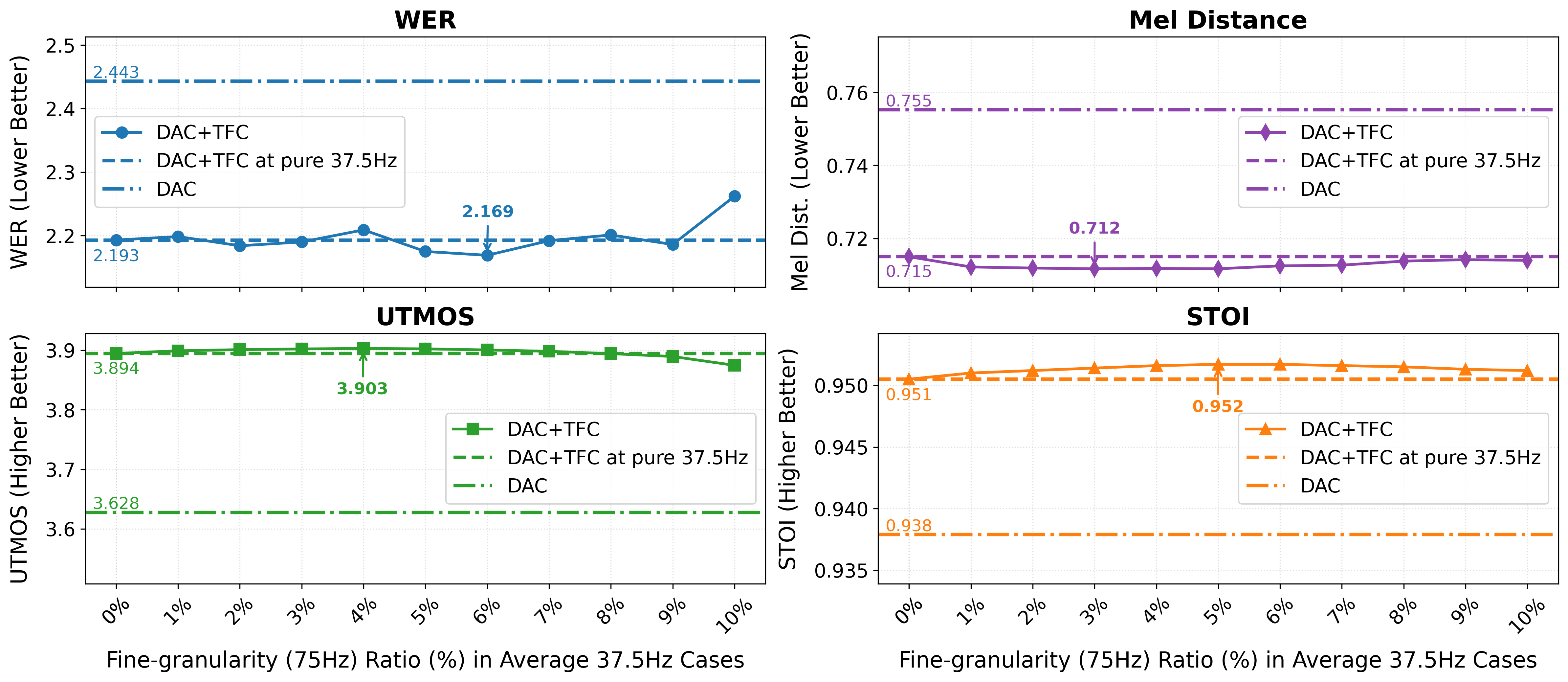}
        \label{average}
    }
    \hfill
    \subfloat[Results for VFR ($<$75Hz) and CFR (75Hz).]{
        \includegraphics[width=0.35\linewidth]{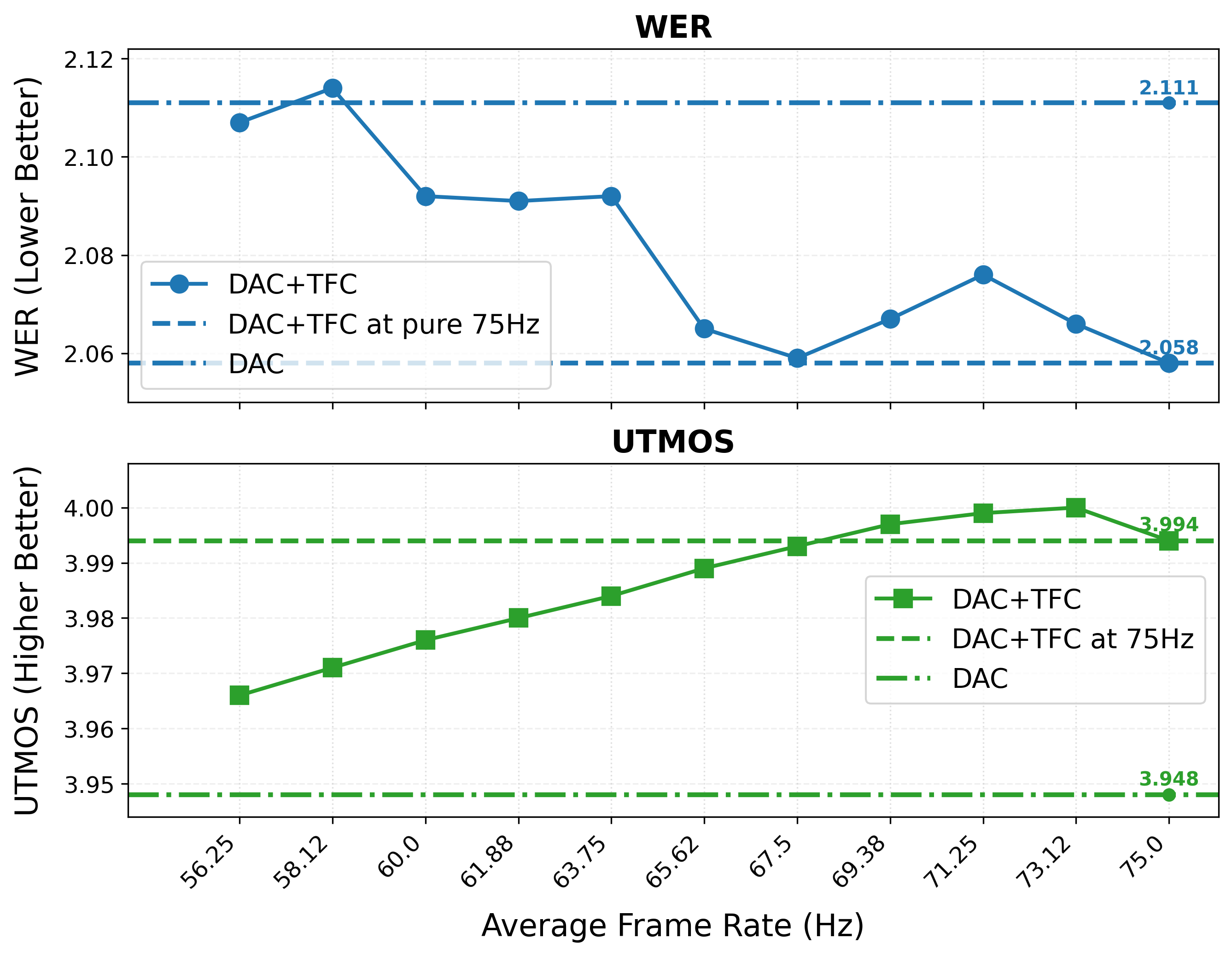}
        \label{smaller}
    }
    \label{fig:two_images}
    \vspace{-0.1in}
    \caption{DAC+TFC (solid lines) is evaluated in VFR mode compared to CFR performances (dashed lines).}
    \vspace{-5pt} % 进一步减少 caption 和正文的距离
\end{figure*}

Both the DAC baseline and the proposed DAC+TFC model are trained on LibriTTS~\cite{libritts}  with 960 hours of speech data. 
To ensure sufficient training across different temporal resolutions, batch inputs for DAC+TFC are heuristically allocated with frame rate ratios $r_f = 0.4,r_m = 0.3,r_c = 0.3$. Each model is trained with a batch size of 20 with 1-second clipped segments for 1000k iterations. 

Reconstruction quality is measured via Mel and STFT distances (configurations following~\cite{dac}), with the latter better capturing high-frequency fidelity. Additionally, perceptual quality is assessed using UTMOS~\cite{utmos}, a MOS prediction system that correlates strongly with human ratings,
% (requiring we downsample the results to 16khz)
 as well as STOI (Short-time Objective Intelligibility) and WER metrics\footnote{Computed with NeMo ASR: \url{https://huggingface.co/nvidia/stt_en_fastconformer_transducer_large}}, ensuring a comprehensive evaluation.

\subsection{Result A: CFR performances across different bitrates}
\label{resultA}
DAC+TFC supports frame rate selection from 18.75Hz to 75Hz by adjusting the granularity ratios in inference.
It can be viewed as an alternative method for bitrate control along with codebook dropout. 
For the DAC baseline, we can select different bitrates by specifying the number of codebooks in inference.
Thus we can conduct fair comparisons between DAC and the proposed DAC+TFC under the same bitrates.
For example, if we use DAC with 4 codebooks to inference, it has a bitrate of 3kbps; then, we compare it with DAC+TFC at 37.5Hz with 8 codebooks that also has a bitrate of 3kbps. 

\begin{table}[h]
\vspace{-5pt}
  \caption{Reconstruction performances of DAC and proposed DAC+TFC in a CFR manner with different bitrates}
  \vspace{-0.1in}
  \setlength{\tabcolsep}{2.4pt}
  \label{tab:eval_results_transposed}
  \centering
  \scriptsize
  \begin{tabular*}{\linewidth}{ccccccccc}
    \toprule
    \makecell{Bitrate\\(kbps)} & Model & 
    \makecell{Frame\\Rate (Hz)} & $N_q$ &
     \makecell{Mel \\ Dist. $\downarrow$} & \makecell{STFT \\ Dist. $\downarrow$} & \makecell{UTMOS \\ $\uparrow$} & \makecell{STOI \\ $\uparrow$} & \makecell{WER \\ $\downarrow$}\\

    \midrule
    \multirow{2}{*}{6} 
    & DAC & 75   &8 & \textbf{0.594}  & 1.456  & 3.948  & 0.967 &4.125 \\
    & DAC+TFC & 75   &8 & 0.601  & \textbf{1.451}  & \textbf{3.994}  & \textbf{0.970} &\textbf{2.845} \\

        \midrule

        \multirow{2}{*}{3} 
    & DAC &75 & 4  & 0.755  & 1.617  & 3.628  & 0.938  &2.443\\
    % & VFR 75Hz 4code   & —       & —       & —       & —       \\
    & DAC+TFC &37.5 &8 & \textbf{0.715}  & \textbf{1.565}  & \textbf{3.895}  & \textbf{0.951}   & \textbf{2.193}\\
    \midrule
    
    \multirow{2}{*}{1.5} 
    & DAC & 75 & 2  & 0.930  & 1.792  & 2.869  & 0.892  &2.110\\
    % & VFR 75Hz 2code   & —       & —       & —       & —       \\
    & DAC+TFC & 18.75 & 8 & \textbf{0.864}  & \textbf{1.725}  & \textbf{2.995}  & \textbf{0.902} & \textbf{2.058} \\
    \bottomrule
  \end{tabular*}
  \vspace{-5pt}
\end{table}
The corresponding results are shown in Table \ref{tab:eval_results_transposed}.
Notably, DAC+TFC achieves better performance than DAC across most metrics, especially in terms of WER.
% Slightly failure in 6kbps' Mel dist. is not important as other metrics have a marginal superior results. 
These results show that TFC effectively controls bitrate, preserves audio quality, and meets the initial goal of reducing sequence length. This can be beneficial for downstream generative models~\cite{audiolm,valle,peng-etal-2024-voicecraft,kyutai2024moshi} since reducing frame count will greatly speed up the generation process, while increased number of codebooks has a negligible impact on generation speed.
% Moreover, even with more codebooks per frame, using a smaller frame rate still offers advantages in downstream generative tasks, since RVQ codes can be modeled by token interleaving or parallel strategy without 

\subsection{Result B: VFR versus CFR performances}
\label{resultB}
In Section \ref{resultA}, while showing basic temporal flexibility, TFC still operates in a CFR way as the frame rates are directly obtained by purely assigning one granularity.
% while true VFR should have at least two no-zero ratio values. 
In this section, we evaluate TFC under truly VFR conditions, i.e. where at least two granularities are combined during inference.

First, we compare VFR and CFR cases at the same average frame rate. 
We adopt an average 3kbps bitrate \& 37.5Hz frame rate setting because it permits mixing components at 18.75Hz and 75Hz (by 2:1 ratios) via TFC to yield VFR results (e.g. by setting $r_f,r_m,r_c$ to $0.2,0.7,0.1$, respectively). 
Figure \ref{average} obviously shows the consistent improvement brought by TFC when mixing other frame rates compared to pure 37.5Hz (0\% 75Hz granularity) and 3kbps DAC in terms of Mel Distance, UTMOS, and STOI metrics. 
Moreover, the lowest WER is achieved with 6\% 75Hz component. 
Overall, these results underscore TFC's flexibility in customizing frame rates and its superior performance benefited from VFR coding.

To further explore the potential of TFC, we compare its VFR results at lower frame rates with 75Hz CFR settings. 
Since lower bitrates from same architecture inevitably degrade reconstruction quality, we focus on UTMOS and WER to assess perceptual quality.
The results (all with \(N_q=8\)) include the 75Hz DAC baseline, as well as DAC+TFC configurations, with frame rate ratios $r_c{:}r_m{:}r_f$ smoothly varying from \(0{:}0{:}100\%\) (purely 75Hz frame rate) to \(0{:}50\%{:}50\%\) (an average of 56.25Hz frame rate). 
As shown in Figure \ref{smaller}, aside from one outlier in WER, the DAC+TFC configurations consistently outperform the 75Hz DAC baseline. 
Notably, some lower frame rate settings even achieve higher UTMOS scores than the 75Hz DAC+TFC configuration, indicating the significant potential of VFR design to compress speeches more compactly.

% \vspace{-0.1in}
\section{Conclusion}

We present Temporally Flexible Coding (TFC), a method that introduces Variable Frame Rate (VFR) into neural speech codecs to dynamically adjust temporal resolution based on information density. TFC effectively balances bitrate and audio quality while reducing sequence length. Due to resource constraints, our implementation is limited to one codec backbone; future work will extend to architecture ablations, as well as evaluations on downstream generation. 
We believe that VFR paradigm holds significant potential for broader applications in neural speech coding and future generative speech models.

\section{Acknowledgements}
This work was supported by the China NSFC Project (No. 92370206), the Shanghai Municipal Science and Technology Major Project (2021SHZDZX0102) and the Key Research and Development Program of Jiangsu Province, China (No.BE2022059).

% \ifinterspeechfinal
%      The Interspeech 2025 organisers
% \else
%      The authors
% \fi

\bibliographystyle{IEEEtran}
\bibliography{mybib}

\end{document}